\documentclass[journal,onecolumn,]{IEEEtran}
\usepackage{enumerate}
\usepackage{xcolor}
\usepackage{cite}
\usepackage{amssymb, amsmath}
\usepackage{amsthm}
\usepackage{graphicx}
\graphicspath{{./figs/}}
\usepackage[caption=false,font=footnotesize]{subfig}
\usepackage{bm}
\usepackage{pifont} 
\usepackage{mathrsfs}

\newcommand{\nop}[1]{}

\newtheorem{definition}{Definition}

\newtheorem{theorem}{Theorem}
\newtheorem{corollary}{Corollary}
\newtheorem{lemma}{Lemma}
\newtheorem{proposition}{Proposition}
\newtheorem{remark}{Remark}
\newtheorem{assumption}{Assumption}

\usepackage{multirow}
\usepackage{tikz}
\usetikzlibrary{arrows}

\begin{document}

\title{Further Properties of Wireless Channel Capacity}
\author{Fengyou~Sun and 
        Yuming~Jiang
}

\maketitle

\begin{abstract}
Future wireless communication calls for exploration of more efficient use of wireless channel capacity to meet the increasing demand on higher data rate and less latency. However, while the ergodic capacity and instantaneous capacity of a wireless channel have been extensively studied, they are in many cases not sufficient for use in assessing if data transmission over the channel meets the quality of service (QoS) requirements. To address this limitation, we advocate a set of wireless channel capacity concepts, namely ``cumulative capacity'', ``maximum cumulative capacity'', ``minimum cumulative capacity'', and ``range of cumulative capacity'', and for each, study its properties by taking into consideration the impact of the underlying dependence structure of the corresponding stochastic process. Specifically, their cumulative distribution function (CDFs) are investigated extensively, where copula is adopted to express the dependence structures. Results considering both generic and specific dependence structures are derived. In particular, in addition to i.i.d., a specially investigated dependence structure is comonotonicity,  i.e, the time series of wireless channel capacity are increasing functions of a common random variable. Appealingly, copula can serve as a unifying technique for obtaining results under various dependence assumptions, e.g. i.i.d. and Markov dependence, which are widely seen in stochastic network calculus. Moreover, some other characterizations of cumulative capacity are also studied, including moment generating function, Mellin transform, and stochastic service curve. With these properties, we believe QoS assessment of data transmission over the channel can be further performed, e.g. by applying analytical techniques and results of the stochastic network calculus theory.
\end{abstract}

\section{Introduction}

In future wireless communication, there will be a continuing wireless data explosion and an increasing demand on higher data rate and less latency. It has been depicted that the amount of IP data handled by wireless networks will exceed $500$ exabytes by 2020, the aggregate data rate and edge rate will increase respectively by $1000\times$ and $100\times$ from 4G to 5G, and the round-trip latency needs to be less than $1$ms in 5G \cite{andrews2014what}.
Evidently, it becomes more and more crucial to explore the ultimate capacity that a wireless channel can provide and to guarantee pluralistic quality of service (QoS) for seamless user experience.

Information theory provides a framework for studying the performance limits in communication and the most basic measure of performance is channel capacity, i.e., the maximum rate of communication for which arbitrarily small error probability can be achieved \cite{tse2005fundamentals}. Due to the time variant nature of a wireless fading channel, its capacity over time is generally a stochastic process. To date, wireless channel capacity has mostly been analyzed for its average rate in the asymptotic regime, i.e., ergodic capacity, or at one time instant/short time slot, i.e., instantaneous capacity. For instance, the first and second order statistical properties of instantaneous capacity have been extensively investigated, e.g. in \cite{rafiq2011statistical,renzo2010channel}.  However, such properties of wireless channel capacity are ordinarily not sufficient for use in assessing if data transmission over the channel meets its QoS requirements. This calls for studying other properties of wireless channel capacity, which can be more easily used for QoS analysis. To meet this need constitutes the objective of this paper. 

Specifically, we advocate in this paper a set of (new) concepts for wireless channel capacity and study their properties. These concepts include ``cumulative capacity'', ``maximum cumulative capacity'', ``minimum cumulative capacity'', and ``range of cumulative capacity''. They respectively refer to the cumulated capacity over a time period, the maximum and the minimum of such capacity within this period, and the gap between the maximum and the minimum. 

Among these (new) concepts, the wireless channel cumulative capacity of a period is essentially the amount of data transmission service that the wireless channel provides (if there is data for transmission) \cite{jiang2005analysis} or is capable of providing (if there is no data for transmission) \cite{jiang2008stochastic} in this period. For the former, the concept is closely related to the (cumulative) service process concept that has been widely used in the stochastic network calculus literature, e.g. in \cite{jiang2005analysis, fidler2006end, jiang2008stochastic, jiang2010note, mahmood2011delay, zheng2013stochastic, ciucu2013towards, ciucu2014sharp, poloczek2015service, leistochastic}. In particular, in these works when charactering the cumulative service process using server models of stochastic network calculus and/or applying the cumulative service process concept to QoS analysis, some special assumptions on the dependence structure of the process are often considered, such as independence \cite{fidler2006end,jiang2008stochastic,jiang2010note} and Markov property \cite{zheng2013stochastic,ciucu2014sharp,poloczek2015service}. 

In addition, we introduce ``maximum cumulative capacity'', ``minimum cumulative capacity'' and ``range of cumulative capacity'' that are new but we believe are also crucial concepts for analyzing QoS performance of wireless channels. This is motivated by the fact that, even with the CDF (i.e. full characteristics) or its bounds of the cumulative capacity known, it may still be difficult to perform QoS analysis of the channel. (One can easily observe this difficulty by assuming fluid traffic input and trying to find backlog bounds from queueing analysis of the channel. See e.g. \cite{jiang2008stochastic}). As a special case of these concepts, forward-looking and backward-looking variations of them are also defined, which turn out to be useful in different application scenarios. 

For the investigation, unlike most existing work in the stochastic network calculus literature, the present paper mainly focuses directly on the cumulative distribution functions (CDFs) of the corresponding processes of these (new) concepts. For their other characterizations, e.g. moment generating function \cite{fidler2006end}, Mellin transform \cite{al2013min}, and stochastic service curve \cite{jiang2008stochastic}, a number of results are also reported for cumulative capacity to exemplify how such properties may be analyzed, but this is not focused. An underlying reason is that a random variable is fully characterized by its CDF. To unify the investigation for each concept, we introduce copula as a technique to account for the various dependence structures implied by different possible properties of the process. For instance, besides a process with i.i.d. increments, a Markov process has also been proved to have a dependence structure \cite{darsow1992copulas,overbeck2015multivariate}. In addition to such dependence assumptions for which many results in the stochastic network calculus are available, we use comonotonicity as a dependence structure when there is a strong time dependence in the channel, i.e., when the time series of instantaneous wireless channel capacities can be represented as increasing functions of a common random variable. Moreover, generic results under arbitrary dependence structures are also obtained when only the marginal distribution functions are assumed to be known. 

To remark, the idea of taking advantage of specific dependence structures in analysis can be found in the stochastic network calculus literature, e.g., independent increments \cite{fidler2006end,jiang2008stochastic,jiang2010note} and Markov property \cite{zheng2013stochastic, ciucu2014sharp,poloczek2015service}. However, such diverse dependence structures are investigated separately, without a unified technique. In addition, the literature investigation mainly focuses on the stochastic service curve characterization of the cumulative service process and on applying it to QoS performance analysis. Little has directly focused on the probability distribution function characteristics of the cumulative capacity. Moreover, to the best of our knowledge, there is no previous work focusing on the probabilistic distribution function characteristics of the maximum, the minimum and the range of cumulative capacity of a wireless channel. In \cite{dong2015copula}, the concept of copula is brought into stochastic network calculus, which is applied to consider the dependence in the superposition property of arrivals. Different from \cite{dong2015copula}, our focus is the cumulative capacity processes that are related to the cumulative service process, while not the arrival process. In addition, we use copula to feature the dependence structures in the considered cumulative capacity processes. 

The contributions of this work\footnote{This is still an on-going work, subject to significant revisions. The contributions are summarized on this on-line version and will be significantly extended when more results are added.} are several-fold. (1) Several concepts for studying wireless channel capacity are introduced, namely ``cumulative capacity'', ``maximum cumulative capacity'', ``minimum cumulative capacity'' and ``range of cumulative capacity''. For the latter three concepts, we originally introduce their forward-looking and backward-looking versions and highlight the fundamental difference between a forward-looking concept and its backward-looking counterpart. (2) A copula technique is introduced to unify the analysis under difference dependence conditions. (3) The probability distribution function characteristics of the processes corresponding to the introduced concepts are investigated, and various exact solutions or bounds on the probability distribution functions are derived. (4) Other characteristics, such as moment generating function, Mellin transform and stochastic service curve, of the cumulative capacity process are investigated to exemplify how such characteristics of the processes corresponding to the introduced concepts can be studied. Based on these properties, QoS assessment of data transmission over a wireless channel can be further performed by, e.g., exploiting the stochastic network calculus theory (see e.g. \cite{jiang2008stochastic}).

The remainder of this paper is structured as follows. The fundamental concepts of wireless channel capacity, including instantaneous capacity, cumulative capacity, maximum cumulative capacity, minimum cumulative capacity and the range of cumulative capacity, are first introduced in Sec. \ref{fundpre}. Also in Sec. \ref{fundpre}, preliminaries for later analysis, including those on copula, non-Granger causality and change of measure, are described. In Sec. \ref{unifying}, copula is elaborated as a unifying technique for analysis of cumulative capacity processes under different dependence structures. The probability distribution function characteristics, particularly the CDF, for cumulative capacity and maximum/minimum cumulative capacity are analyzed respectively in Sec. \ref{analysiscc} and Sec. \ref{analysismcc}. Other characterizations of cumulative capacity are elaborated in Sec. \ref{othercht}. Finally, the paper is concluded and future work is discussed in Sec. \ref{conclusion}.

\section{Fundamental Concepts and Preliminaries}\label{fundpre}

\subsection{Fundamental Concepts}

\subsubsection{Instantaneous Capacity}

Consider a wireless channel. We assume discrete time $t= 1, 2, \dots$, and that the instantaneous capacity \cite{costa2010multiple} or mutual information \cite{telatar1999capacity} $C(t)$ of the channel at time $t$ can be expressed as a function of the instantaneous SNR $\gamma_t$ at this time \cite{rafiq2011statistical}:
\begin{equation}
C(t)=\log_{2}(g(\gamma_t)).
\end{equation}

For single input single output (SISO) channels, 
if CSI is only known at the receiver,
the instantaneous capacity or the mutual information of the channel, assuming flat fading, can be expressed as
\begin{equation}
C(t)=\log_{2}(1+\gamma|h(t)|^{2}),
\end{equation}
where $h(t)$ is a stochastic process describing the fading behavior, $|h(t)|$ denotes the envelope of $h(t)$, $\gamma={P}/{N_{0}W}$ denotes the average received SNR per complex degree of freedom, $P$ is the average transmission power per complex symbol, $N_{0}/2$ is the power spectral density of AWGN, and $W$ is the channel bandwidth. 

In the literature, PDF or CDF of the instantaneous capacity is available for various types of channels, e.g. Rayleigh channel \cite{hogstad2007exact}, Rice channel \cite{rafiq2009statisticalcomb}, Nakagami-$m$ channel \cite{rafiq2008statisticalmimo}, Suzuki channel \cite{rafiq2007impact}, and more \cite{rafiq2011statistical}.
Specifically, the CDF of the Rayleigh channel instantaneous capacity is expressed as \cite{hogstad2007exact}
\begin{eqnarray}
F_{C(t)}(r) &=& 1-e^{-(2^r-1)/\gamma}. \label{rayleighf}
\end{eqnarray}

\subsubsection{Cumulative Capacity}  

We define the {\em cumulative capacity} through period $(s,t]$ as 
\begin{equation}
S(s,t) \equiv \sum\limits_{i=s+1}\limits^{t}{C(i)},
\end{equation} 
where $C(i)$ is the instantaneous capacity at time $i$.

\subsubsection{Maximum Cumulative Capacity}

We define the {\em maximum cumulative capacity} in period $(0,t]$ as 
\begin{equation}\label{g-MCC}
\overline{S}(0,t) \equiv \sup_{{1}\le{j}\le{k}\le{t}}S(j,k) = \sup_{{1}\le{j}\le{k}\le{t}}\left(\sum\limits_{i=j}\limits^{k}{C(i)}\right),
\end{equation} 
where $C(i)$ is the instantaneous capacity at time $i$. 

Fixing $j=1$ in (\ref{g-MCC}), we obtain a {\em forward-looking} of the maximum cumulative capacity, i.e.,
\begin{equation} 
\overrightarrow{S}(0,t) \equiv \sup_{1\le k\le t} \overline{S}(0,k) = \sup_{{1}\le{k}\le{t}}\left(\sum\limits_{i=1}\limits^{k}{C(i)}\right),
\end{equation}
while fixing $k=t$ in (\ref{g-MCC}), we obtain a {\em backward-looking} of the maximum cumulative capacity, i.e.,
\begin{equation} 
\overleftarrow{S}(0,t) \equiv \sup_{1\le j\le t} \overline{S}(j,t) = \sup_{{1}\le{j}\le{t}}\left(\sum\limits_{i=j}\limits^{t}{C(i)}\right).
\end{equation}

\nop{
We define the {\em maximum cumulative capacity} through period $(s,t]$ as 
\begin{equation}
\overline{S}(s,t) \equiv \sup_{{s+1}\le{j}\le{k}\le{t}}S(j,k) = \sup_{{s+1}\le{j}\le{k}\le{t}}\left(\sum\limits_{i=j}\limits^{k}{C(i)}\right),
\end{equation} 
where $C(i)$ is the instantaneous capacity at time $i$. 
Fix $j=s+1$, we obtain a {\em forward-looking} of the maximum cumulative capacity, i.e.,
\begin{equation} 
\overrightarrow{S}(s,t) \equiv \sup_{s+1\le i\le t} \overline{S}(s,i) = \sup_{{s+1}\le{k}\le{t}}\left(\sum\limits_{i=s+1}\limits^{k}{C(i)}\right),
\end{equation}
while fix $k=t$, we obtain a {\em backward-looking} of the maximum cumulative capacity, i.e.,
\begin{equation} 
\overleftarrow{S}(s,t) \equiv \sup_{s+1\le i\le t} \overline{S}(i,t) = \sup_{{s+1}\le{k}\le{t}}\left(\sum\limits_{i=k}\limits^{t}{C(i)}\right).
\end{equation}
}

\subsubsection{Minimum Cumulative Capacity}

We define the {\em minimum cumulative capacity} in period $(0,t]$ as 
\begin{equation} \label{g-mcc}
\underline{S}(0,t) \equiv \inf_{{1}\le{j}\le{k}\le{t}}S(j,k) = \inf_{{1}\le{j}\le{k}\le{t}}\left(\sum\limits_{i=j}\limits^{k}{C(i)}\right),
\end{equation} 
where $C(i)$ is the instantaneous capacity at time $i$.

Similarly, fixing $j=1$ in (\ref{g-mcc}), we obtain a {\em forward-looking} of the minimum cumulative capacity, i.e.,
\begin{equation} 
\underrightarrow{S}(0,t) \equiv \inf_{1\le k\le t} \underline{S}(0,k) = \inf_{{1}\le{k}\le{t}}\left(\sum\limits_{i=1}\limits^{k}{C(i)}\right),
\end{equation}
and fixing $k=t$ in (\ref{g-mcc}), we obtain a {\em backward-looking} of the minimum cumulative capacity, i.e.,
\begin{equation} 
\underleftarrow{S}(0,t) \equiv \inf_{1\le j\le t} \underline{S}(j,t) = \inf_{{1}\le{j}\le{t}}\left(\sum\limits_{i=j}\limits^{t}{C(i)}\right).
\end{equation}

\nop{
We define the {\em minimum cumulative capacity} through period $(s,t]$ as 
\begin{equation}
\underline{S}(s,t) \equiv \inf_{{s+1}\le{j}\le{k}\le{t}}S(j,k) = \inf_{{s+1}\le{j}\le{k}\le{t}}\left(\sum\limits_{i=j}\limits^{k}{C(i)}\right),
\end{equation} 
where $C(i)$ is the instantaneous capacity at time $i$.
Fix $j=s+1$, we obtain a {\em forward-looking} of the minimum cumulative capacity, i.e.,
\begin{equation} 
\underrightarrow{S}(s,t) \equiv \inf_{s+1\le i\le t} \underline{S}(s,i) = \inf_{{s+1}\le{k}\le{t}}\left(\sum\limits_{i=s+1}\limits^{k}{C(i)}\right),
\end{equation}
while fix $k=t$, we obtain a {\em backward-looking} of the minimum cumulative capacity, i.e.,
\begin{equation} 
\underleftarrow{S}(s,t) \equiv \inf_{s+1\le i\le t} \underline{S}(i,t) = \inf_{{s+1}\le{k}\le{t}}\left(\sum\limits_{i=k}\limits^{t}{C(i)}\right).
\end{equation}
}

\subsubsection{Range of Cumulative Capacity}
We define the {\em range of cumulative capacity} in period $(0,t]$ as
\begin{equation}
R(0,t) \equiv \overline{S}(0,t) - \underline{S}(0,t).
\end{equation}
The range can also have variations based on the selection of forward-looking and backward-looking expressions of the maximum and minimum cumulative capacity, e.g, forward-looking $\overrightarrow{R}(0,t) \equiv \overrightarrow{S}(0,t) - \overrightarrow{S}(0,t)$ and backward-looking $\underrightarrow{R}(0,t) \equiv \underrightarrow{S}(0,t) - \underrightarrow{S}(0,t)$

\nop{
We define the {\em range of cumulative capacity} through period $(s,t]$ as
\begin{equation}
R(s,t) \equiv \overline{S}(s,t) - \underline{S}(s,t).
\end{equation}
The range can also have a series of variations based on the selection of forward-looking and backward-looking expressions of the maximum and minimum cumulative capacity, e.g, $\overrightarrow{R}(s,t) = \overrightarrow{S}(s,t) - \underrightarrow{S}(s,t)$.
}

\begin{remark}
$S(0,t)$ is essentially a partial sum process in probability theory.  When studying the maxima of partial sums in probability theory, the forward-looking versions of $\overline{S}(0,t) $ and $\overline{R}(0,t)$ are typically focused \cite{gomide1975range,karlin1981second}. This is probably due to that the studied probability problems are often forward-looking in nature, e.g., at time $0$, making decision based on possible happenings in the future time $t$. However, for QoS analysis of communication networks, backward-looking is more important, e.g. when deciding delay at time $t$, we have to look backwards to see how much service the follow has experienced. As directly seen from their definitions, the forward-looking maximum cumulative capacity is fundamentally different from its backward-looking counterpart. As a consequence, results in probability theory for the process of maxima of partial sums should be used with care when they are applied to queueing analysis. In general, under time reversibility assumption, the results for the forward-looking definitions may be extended for application to the backward-looking definitions.

The range describes the gap between the maximum and the minimum, and hence the smaller the range, the closer are the two. In particular, when the range is small or approaching $0$, the maximum becomes the minimum, hence should also equal the cumulative capacity. Range may be used as a measure to characterize the tightness between an upper bound and a lower bound on the cumulative capacity. The idea of using a gap between analytical bounds themselves or between the bounds and the exact results to characterize the tightness or accuracy of the obtained bounds has been used in e.g. Information Theory to investigate / characterize the information capacity of a channel \cite{cover2006elements}. In the context of stochastic network calculus, such an idea was exploited in \cite{LuYJ14} to study the accuracy of the obtained bounds.


\end{remark}

\subsection{Preliminaries}

\subsubsection{Copula}

Copula is a well-known concept for dependence modeling by decoupling the joint distribution function into the dependence structure and marginal behavior.

\begin{definition} \cite{embrechts2005quantitative,nelsen2006introduction}
A $d$-dimensional copula is a distribution function on $[0, 1]^d$ with standard uniform marginal distributions.
\end{definition}

It is equivalent to say that a copula is any function $C:[0,1]^n\rightarrow[0,1]$, which has the following three properties: \cite{embrechts2005quantitative,nelsen2006introduction}
\begin{description}
\item (1)
$C(u_1,\ldots,u_d)$ is increasing in each component $u_i$.
\item (2)
$C(1,\ldots,1,u_i,1,\ldots,1)=u_i$ for all $i\in\{1,\ldots,d\}$, $u_i\in[0,1]$.
\item (3)
For all $(a_1,\ldots,a_d), (b_1,\ldots,b_d) \in[0,1]^d$ with $a_i\le{b_i}$, 
$\sum_{i_1=1}^{2}\ldots\sum_{i_d=1}^{2}(-1)^{i_1+\ldots+i_d}C(u_{1,i_1},\ldots,u_{d,i_d})\ge{0}$, 
where $u_{j,1}=a_j$ and $u_{j,2}=b_j$ for all $j\in\{1,\ldots,d\}$.
\end{description}

The significance of copulas in studying the joint distribution
functions is summarized by the Sklar's theorem, which shows that all joint distribution functions contain copulas and copulas may be used in conjunction with marginal distribution functions to construct joint distribution functions.

\begin{theorem} \cite{embrechts2005quantitative}
Let $F$ be a joint distribution function with marginals $F_1,\ldots,F_d$. Then there exists a copula $C: [0,1]^d\rightarrow[0,1]$ such that, for all $x_1,\ldots,x_d$ in $\bar{\mathbb{R}}=[-\infty,\infty]$
\begin{equation}
F(x_1,\ldots,x_d)=C(F_1(x_1),\ldots,F_d(x_d)). 
\end{equation}
If the marginals are continuous, then $C$ is unique; otherwise $C$ is uniquely determined on $Ran{F_1}\times{Ran}F_2\times\ldots\times{Ran}F_d$, where $Ran{F_i}=F_i(\bar{\mathbb{R}})$ denotes the range of $F_i$. Conversely, if $C$ is a copula and $F_1,\ldots,F_d$ are univariate distribution functions, then the function $F$ is a joint distribution function with marginals $F_1,\ldots,F_d$.
\end{theorem}

\subsubsection{Non-Granger Causality}

Non-Granger causality is a concept initially introduced in econometrics and refers to a multivariate dynamic system in which each variable is determined by its own lagged values and no further information is provided by the lagged values of the other variables. This concept has a direct copula expression.
\begin{assumption}\label{nongranger} \cite{cherubini2010dependence}
The copula function
\begin{equation}
C^{t_j}\left(F_{X_1}^{t_1},F_{X_1}^{t_2},\ldots,F_{X_1}^{t_j},\ldots,F_{X_m}^{t_1},F_{X_m}^{t_2},\ldots,F_{X_m}^{t_j}\right)
\end{equation}
admits the hierarchical representation
\begin{equation}
C^{t_j}\left(G_{X_1}^{t_j}\left(F_{X_1}^{t_1},F_{X_1}^{t_2},\ldots,F_{X_1}^{t_j}\right),\ldots,G_{X_m}^{t_j}\left(F_{X_m}^{t_1},F_{X_m}^{t_2},\ldots,F_{X_m}^{t_j}\right)\right),
\end{equation}
where $G_{X_i}^{t_j}(u_i^1,u_i^2,\ldots,u_i^j)$, $i=1,2,\ldots,m$ are copula functions.
\end{assumption}

Denote the running minimum and maximum of $X_i$ up to time $t_j$ respectively as
\begin{eqnarray}
m_i(t_j) &=& \min\{ X_i(t_k); t_1\le t_k\le t_j \}, \\
M_i(t_j) &=& \max\{ X_i(t_k); t_1\le t_k\le t_j \}
\end{eqnarray}
and define 
\begin{eqnarray}
\overline{F}_{m_i}^{t_j}(B_i) &=& P(m_i(t_j)> B_i),\\ 
F_{M_i}^{t_j}(B_i) &=& P(M_i(t_j)\le B_i).
\end{eqnarray}
\begin{proposition} \cite{cherubini2010dependence}
Assume 
\begin{equation}
P(X_1(t_n)\le B_1,\ldots,X_k(t_n)\le B_k) = C^{t_n}\left(F_{X_1}^{t_n}(B_1),\ldots,F_{X_k}^{t_n}(B_k)\right),
\end{equation}
and that the copula function allows the hierarchical representation under Assumption \ref{nongranger}. Then the copula function representing the dependence structure among the running maximum (minimum) at time $t_n$ is the same copula function (survival copula function) representing dependence among the levels at the same time, namely,
\begin{eqnarray}
P(M_1(t_n)\le B_1,\ldots,M_k(t_n)\le B_k) &=& {C}^{t_n}\left({F}_{M_1}^{t_n}(B_1),\ldots,{F}_{M_k}^{t_n}(B_k)\right), \\
P(m_1(t_n)> B_1,\ldots,m_k(t_n)> B_k) &=& \overline{C}^{t_n}\left(\overline{F}_{m_1}^{t_n}(B_1),\ldots,\overline{F}_{m_k}^{t_n}(B_k)\right).
\end{eqnarray}
\end{proposition}

\subsubsection{Change of Measure}

Consider a stochastic processes $\{Z_t\}$ with a Polish state space $E$ and
sample paths in the Skorokhod space $D=D\left([0,\infty),E\right)$ equipped with the natural filtration $\{\mathscr{F}_t\}_{t\ge{0}}$ and the Borel $\sigma$-field $\mathscr{F}$.
For two processes represented by probability measures $\mathbb{P}$, $\widetilde{\mathbb{P}}$ on $(D,\mathscr{F})$, it is interesting to look for
a likelihood ratio process $\{L_t\}$, such that
\begin{eqnarray}
\widetilde{\mathbb{P}} = \mathbb{E}[L_t,A], ~A\in\mathscr{F}_t, \label{likelirp}
\end{eqnarray}
i.e., the restriction of $\widetilde{\mathbb{P}}$ to $(D,\mathscr{F}_t)$ is absolutely continuous w.r.t. the restriction of $\mathbb{P}$ to $(D,\mathscr{F}_t)$ \cite{asmussen2003applied}\cite{piccioni1983continuous}.

\begin{proposition} \cite{asmussen2003applied} \label{lhrpcm}
Let $\{\mathscr{F}_t\}_{t\ge{0}}$ be the natural filtration on $D$, $\mathscr{F}$ the Borel $\sigma$-field and $\mathbb{P}$ a given probability measure on $(D,\mathscr{F})$.
\begin{description}
\item{(i)} 
If $\{L_t\}_{t\ge{0}}$ is a nonnegative martingale w.r.t. $(\{\mathscr{F}_t\},\mathbb{P})$ such that $\mathbb{E}{L_t}=1$, then there exists a unique probability measure $\widetilde{\mathbb{P}}$ on $\mathscr{F}$ such that (\ref{likelirp}) holds.
\item{(ii)} 
Conversely, if for some probability measure $\widetilde{\mathbb{P}}$ and some $\mathscr{F}_t$-adapted process $\{L_t\}_{t\ge{0}}$ (\ref{likelirp}) holds, then $\{L_t\}$ is a nonnegative martingale w.r.t. $(\{\mathscr{F}_t\},\mathbb{P})$ such that $\mathbb{E}{L_t}=1$.
\end{description}
\end{proposition}

\begin{theorem} \cite{asmussen2003applied}
Let $\{L_t\}$, $\widetilde{\mathbb{P}}$ be as in Proposition \ref{lhrpcm}(i). If $\tau$ is a stopping time and $G\in\mathscr{F}_\tau$, $G\subseteq\{ \tau<\infty \}$, then
\begin{equation}
\mathbb{P}(G) = \widetilde{\mathbb{E}}\left[ \frac{1}{L_\tau}; G \right].
\end{equation}
More generally, if the waiting time process $W\ge{0}$ is $\mathscr{F}_\tau$-measurable, then $\mathbb{\mathbb{E}}[W; \tau<\infty]=\widetilde{\mathbb{E}}[W/L_\tau;\tau<\infty]$.
\end{theorem}

\begin{corollary} \cite{asmussen2003applied}
Let $\{L_t\}$, $\widetilde{\mathbb{P}}$ be as in Proposition \ref{lhrpcm}(i), and let $\tau$ be a stopping time with $\mathbb{P}(\tau<\infty)=1$. Then a necessary and sufficient condition that $\mathbb{E}L_\tau=1$ is that $\widetilde{\mathbb{P}}(\tau<\infty)=1$.
\end{corollary}

\begin{definition} \cite{asmussen2003applied}
Assume that $\{Z_t\}$ is Markov w.r.t. the natural filtration ${\mathscr{F}_t}$ on $D$ and define $\{L_t\}$ to be a multiplicative functional if $\{Lt\}$ is adapted to ${\mathscr{F}_t}$ and
\begin{eqnarray}
L_{t+s} = L_{t}\cdot(L_{s}\circ\theta_{t})
\end{eqnarray}
$\mathbb{P}_x$-a.s. for all $x$, $s$, $t$, where $\theta_t$ is the shift operator. The precise meaning of this is the following: being $\mathscr{F}_t$-measurable, $L_t$ has the form $L_t=\varphi_{t}(\{Z_u\}_{0\le{u}\le{t}})$ for some mapping $\varphi_t: D[0,t]\rightarrow[0,\infty)$, and then $L_s\circ\theta_t=\varphi_s(\{Z_{t+u}\}_{0\le{u}\le{s}})$.
\end{definition}

\begin{theorem} \cite{asmussen2003applied} \label{makovprop}
Let $\{Z_t\}$ be Markov w.r.t. the natural filtration $\{\mathscr{F}_t\}$ on
$D$, let $\{L_t\}$ be a nonnegative martingale with $\mathbb{E}_x{L_t} = 1$ for all $x$, $t$ and let $\widetilde{P}_x$ be the probability measure given by $\widetilde{P}_x(A)=\mathbb{E}_x[L_t;A]$. Then the family $\{\widetilde{P}_x\}_{x\in{E}}$ defines a time-homogeneous Markov process if and only if $\{L_t\}$ is a multiplicative functional. A multiplicative functional $\{L_t\}$ with $\mathbb{E}_x{L_t} = 1$ for all $x$, $t$ is a martingale.
\end{theorem}

A Markov additive process is defined as a bivariate Markov process $\{X_t\}=\{(J_t,S_t)\}$ where $\{J_t\}$ is a Markov process with state space $E$ and the increments of $\{S_t\}$ are govenrened by $\{J_t\}$ in the sense that 
\begin{equation}
\mathbb{E}[f(S_{t+s}-S_t)g(J_{t+s})|\mathscr{F}_t] = \mathbb{E}_{J_t,0}[f(S_s)g(J_s)].
\end{equation}
In discrete time, a Markov additive process is specified by the measure-valued matrix (kernel) $\mathbf{F}(dx)$ whose $ij$th element is the defective probability distribution 
\begin{equation}
F_{ij}(dx)=\mathbb{P}_{i,0}(J_1=j,Y_1\in{dx}),
\end{equation}
where $Y_n=S_n-S_{n-1}$. An alternative description is in terms of the transition matrix $\mathbf{P}=(p_{ij})_{i,j\in{E}}$ (here $p_{ij}=\mathbb{P}_i(J_1=j)$) and the probability measures
\begin{equation}
H_{ij}(dx) = \mathbb{P}(Y_1\in{dx}|J_0=i, J_1=j) = \frac{F_{ij}(dx)}{p_{ij}}.
\end{equation} 
We denote the $E\times{E}$ matrix $\widehat{\textbf{F}}[\theta]$ with $ij$the element $\widehat{F}^{(ij)}[\theta]=:\int{e^{\theta{x}}F^{(ij)}}(dx)$. By Perron-Frobenius theory, the matrix $\widehat{\textbf{F}}[\theta]$ has a positive real eigenvalue with maximal absolute value $e^{\kappa(\theta)}$ and the corresponding right eigenvector $\textbf{h}^{(\theta)}=(h_{i}^{(\theta)})_{i\in{E}}$, i.e., $\widehat{\textbf{F}}[\theta]\textbf{h}^{(\theta)}=e^{\kappa(\theta)}\textbf{h}^{(\theta)}$.
The exponential change of measure corresponding to $\theta$ is then given by
\begin{eqnarray}
\widetilde{\mathbf{P}} &=& e^{-\kappa(\theta)}\mathbf{\Delta}^{-1}_{\mathbf{h}^{(\theta)}}\widehat{\textbf{F}}[\theta]\Delta_{\mathbf{h}^{(\theta)}}, \\
\widetilde{H}_{ij}(dx) &=& \frac{e^{\theta{x}}}{\widehat{H}_{ij}[\theta]}H_{ij}(dx),
\end{eqnarray}
where $\mathbf{\Delta}_{\mathbf{h}^{(\theta)}}$ is the diagonal matrix with the $h_{i}^{(\theta)}e^{\theta{x}}$ on the diagonal, in particular, $\widetilde{p}_{ij}=e^{-\kappa(\theta)}p_{ij}h_{j}^{(\theta)}/h_{i}^{\theta}$, and $\widehat{H}_{ij}[\theta]$ is the normalizing constant.
The likelihood ratio is 
\begin{equation}
L_n = \frac{h^{(\theta)}(J_n)}{h^{(\theta)}(J_0)}e^{-\theta{S_n}+n\kappa(\theta)},
\end{equation}
which is a mean-one martingale \cite{asmussen2003applied}.

\section{Copula: A Unifying Technique for Analysis}\label{unifying}

\subsection{Bounds of Dependence Structures}

For a random vector $\textbf{X}=(X_1, \ldots, X_n)$, where the marginal distribution functions $F_i \sim X_i$ are known but the dependence between the components is unspecified. Denote the sharp lower and upper Fr\'echet bounds over all dependence structures as
$M_n(t) := \sup \left\{ P\left( \sum_{i=1}^{n}X_i\le t \right); X_i\sim F_i, 1\le{i}\le{n} \right\}$, $m_n(t) := \inf \left\{ P\left( \sum_{i=1}^{n}X_i< t \right); X_i\sim F_i, 1\le{i}\le{n} \right\}$, and denote $M_{n}^{+}(t) := 1-m_n(t)$, $m_{n}^{+}(t) := 1-M_n(t)$.
The sharp Fr\'echet bounds for the case $n=2$ were derived in \cite{makarov1982estimates,ruschendorf1982random}, and has been extended to $n\ge 3$, namely the standard bounds \cite{frank1987best,puccetti2012bounds,ruschendorf2013mathematical}.

\nop{
Let
\begin{eqnarray}
M_n(t) &:=& \sup \left\{ P\left( \sum_{i=1}^{n}X_i\le t \right); X_i\sim F_i, 1\le{i}\le{n} \right\}, \\
m_n(t) &:=& \inf \left\{ P\left( \sum_{i=1}^{n}X_i< t \right); X_i\sim F_i, 1\le{i}\le{n} \right\}.
\end{eqnarray}
Then 
\begin{eqnarray}
M_{n}^{+}(t) &:=& 1-m_n(t)=\sup \left\{ P\left( \sum_{i=1}^{n}X_i\ge t \right); X_i\sim F_i, 1\le{i}\le{n} \right\},\\ 
m_{n}^{+}(t) &:=& 1-M_n(t)=\inf \left\{ P\left( \sum_{i=1}^{n}X_i> t \right); X_i\sim F_i, 1\le{i}\le{n} \right\}.
\end{eqnarray}
}

\subsubsection{Standard Bounds}

\begin{theorem}\label{copula} \cite{frank1987best,puccetti2012bounds,ruschendorf2013mathematical}
Let $X_i\sim F_i$, $1\le i\le d$. Then, for any $s\in\mathbb{R}$, we have that
\begin{eqnarray}
\max\left\{ \sup_{\mathbf{u}\in\mathcal{U}(s)}\left\{ \sum_{i=1}^{d}F_i(u_i) \right\}-(d-1), 0 \right\} \le {P}\left(\sum_{i=1}^{d}X_i\le s \right) \le \min\left\{ \inf_{\mathbf{u}\in\mathcal{U}(s)}\left\{ \sum_{i=1}^{d}F_i(u_i) \right\}, 1 \right\},
\end{eqnarray}
where $\mathcal{U}(s)=\left\{ \mathbf{u}=(u_1,\ldots,u_d)\in\mathbb{R}^d:\sum_{i=1}^{d}u_i=s \right\}$.
\end{theorem}

\begin{remark}
For the fast computation of standard bounds, a numerical method is described
in \cite{embrechts2003using}, while an analytical method is described in \cite{embrechts2006aggregating}.
\end{remark}

\subsubsection{Dual Bounds}

The standard bounds are not sharp for $n\ge 3$ and an improvement can be obtained based on duality theorems. Specifically, $M_{n}^{+}(t)$ and $m_{n}^{+}(t)$ have the following dual counter parts \cite{puccetti2012bounds,ruschendorf2013mathematical}
\begin{eqnarray}
M_{n}^{+}(s) &=& \inf \left\{ \sum_{i=1}^{n}\int{g_i}d{F_i}; g_i \text{~bounded}, 1\le{i}\le{n} \text{~with} \sum_{i=1}^{n}g_i(x_i)\ge 1_{[s,+\infty)}\left( \sum_{i=1}^{n}X_i \right) \right\},\\
m_{n}^{+}(s) &=& \sup \left\{ \sum_{i=1}^{n}\int{f_i}d{F_i}; f_i \text{~bounded}, 1\le{i}\le{n} \text{~with} \sum_{i=1}^{n}f_i(x_i)\le 1_{[s,+\infty)}\left( \sum_{i=1}^{n}X_i \right) \right\}.
\end{eqnarray}

While the dual representations are difficult to evaluate in general, they allow to establish good bounds obtained by choosing admissible piecewise linear dual functions in the dual problem.

\begin{theorem}\label{dualbasic} \cite{puccetti2012bounds,ruschendorf2013mathematical}
Let $X_i\sim F_i$ and $\overline{F_i} = 1-F_i$ be the survival function of $F_i$. Then, for any $s\in\mathbb{R}$, we have 
\begin{eqnarray}
M_{n}^{+}(s) &\le& D(s) = \inf_{u\in\overline{\mathcal{U}}(s)} \min\left\{ \frac{\sum_{i=1}^{n}\int_{u_i}^{s-\sum_{j\neq{i}}u_j}\overline{F_i}(t)dt}{s-\sum_{i=1}^{n}u_i}, 1 \right\},\\
m_{n}^{+}(s) &\ge& d(s) = \sup_{u\in\underline{\mathcal{U}}(s)} \max\left\{ \frac{\sum_{i=1}^{n}\int_{u_i}^{s-\sum_{j\neq{i}}u_j}\overline{F_i}(t)dt}{s-\sum_{i=1}^{n}u_i}-n+1, 0 \right\},
\end{eqnarray}
where $\overline{\mathcal{U}}(s) = \left\{ u\in\mathbb{R}^{n}; \sum_{i=1}^{n}u_i <s \right\}$ and $\underline{\mathcal{U}}(s) = \left\{ u\in\mathbb{R}^{n}; \sum_{i=1}^{n}u_i >s \right\}$.
\end{theorem}

Let $u\in\mathcal{U}(s)$, $\sum_{i=1}^{n}u_i=s$, the piecewise linear dual admissible choices become piecewise constant and thus yield a standard bound. As a consequence, the dual bounds improve the corresponding standard bounds \cite{puccetti2012bounds,ruschendorf2013mathematical}.
However, the calculation of the dual bounds requires to solve an $n$-dimensional optimization problem which typically will be possible only for small values of $n$. For the homogeneous case, $F_i=F$, $1\le{i}\le{n}$, a simplified expression can be obtained with a one-dimensional problem that can be solved in any dimension.

\begin{theorem} \cite{puccetti2012bounds}
Let $F_1=\ldots=F_n =: F$ be distribution functions on $\mathbb{R}_{+}$. Then for any $s\ge{0}$ it holds that
\begin{eqnarray}
M_{n}^{+}(s) &\le& D(s) = \inf_{u<{s/n}}\min\left\{ \frac{n\int_{u}^{s-(n-1)u}\overline{F}(t)dt}{s-nu}, 1 \right\}, \\
m_{n}^{+}(s) &\ge& d(s) = \sup_{u>{s/n}}\max\left\{ \frac{n\int_{u}^{s-(n-1)u}\overline{F}(t)dt}{s-nu} -n+1, 0 \right\}.
\end{eqnarray}
\end{theorem}

\begin{theorem} \cite{embrechts2006bounds}
Let $F_1=\ldots=F_n =: F$ be distribution functions on $\mathbb{R}_{+}$. Then for any $s\ge{0}$ it holds that
\begin{equation}
m_{n}(t)\ge 1- n \inf_{r\in[0,s/n)}\frac{\int_{r}^{s-(n-1)r}\overline{F}(x)dx}{s-nr}. \label{sharp_bound}
\end{equation}
\end{theorem}

The infimum in (\ref{sharp_bound}) can be easily calculated numerically by finding the zero derivative points of its argument in the specified interval. Note that 
\begin{equation}
\lim_{r\rightarrow{s/n}}\left\{ 1- n\frac{\int_{r}^{s-(n-1)r}\overline{F}(x)dx}{s-nr} \right\} = nF(s/n) - n +1,
\end{equation}
which means that (\ref{sharp_bound}) is greater or equal than the standard lower bound. Moreover, the dual bounds in (\ref{sharp_bound}) have been proved to be sharp under some general distributional assumptions \cite{puccetti2013sharp}:
\begin{description}
\item{(A1) Attainment condition:}
There exists some $a<s/n$ such that
\begin{equation}
D(s) = \inf_{t<s/n}\frac{n\int_{t}^{s-(n-1)t}\overline{F}(x)dx}{s-nt} = \frac{n\int_{a}^{b}\overline{F}(x)dx}{b-a},
\end{equation}
where $b=s-(n-1)a$ and $a^{\ast}=F^{-1}(1-D(s))\le{a}$.
\item{(A2) Mixability condition:}
The conditional distribution of $(X_1 | X_1\ge{a^\ast})$ is $n$-mixable on $(a,b)$.
\item{(A3) Ordering condition:}
For all $y\ge{b}$ it holds that
\begin{equation}
(n-1)(F(y)-F(b))\le F(a) - F\left( \frac{s-y}{n-1} \right).
\end{equation}
\end{description}

\begin{theorem}\label{dualsharp} \cite{puccetti2013sharp}
Under the attainment condition (A1), the mixing condition (A2) and the ordering condition (A3), the dual bound is sharp, that is 
\begin{equation}
M_{n}^{+}(s) = D(s) = \inf_{t<s/n}\frac{n\int_{t}^{s-(n-1)t}\overline{F}(x)dx}{s-nt} = \frac{n\int_{a}^{b}\overline{F}(x)dx}{b-a}.
\end{equation}
\end{theorem}

The first order conditions for the optimization in the attainment assumption (A1) at $t=a$ imply that 
\begin{equation}
\frac{n\int_{a}^{b}\overline{F}(x)dx}{b-a} = \overline{F}(a)+(n-1)\overline{F}(b),
\end{equation}
where $b=s-(n-1)a$ and $a^{\ast}=F^{-1}(1-D(s))\le{a}$, and it provides a clue to calculate the basic point $a$ and, hence, the dual bound $D(s)$. Having calculated $a$ one can easily check the second order condition
\begin{equation}
f(a)-(n-1)^{2}f(b) \ge{0},
\end{equation}
which is necessary to guarantee that $a$ is a point of minimum for (A1).
At this point, the sharpness of the dual bound $D(s)$ can be obtained from a different set of assumptions \cite{puccetti2013sharp}:
\begin{enumerate}
\item
Continuous distribution functions $F$ having a positive and decreasing density $f$ on $(a^\ast,\infty)$ satisfy assumptions (A2) and (A3).
\item
Continuous distribution functions $F$ having a concave
density $f$ on the interval $(a,b)$ satisfy the mixing assumption (A2). In order to obtain sharpness of the dual bound $D(s)$ for these distributions, conditions (A1) and (A3) have to be checked numerically.
\end{enumerate}

\subsection{Example Dependence Structures}

\subsubsection{Comonotonicity}

\begin{definition} \cite{dhaene2002concept}
The set $A\subseteq\mathbb{R}^n$ is said to be comonotonic if for any $\underline{x}\le\underline{y}$ or $\underline{y}\le\underline{x}$ holds, where $\underline{x}\le\underline{y}$ denotes the componentwise order, i.e., $x_i\le y_i$ for all $i=1,2,\ldots,n$. 
\end{definition}

\begin{definition} \cite{dhaene2002concept}
A random vector $\underline{X}=(X_1,\ldots,X_n)$ is said to be comonotonic it has a comonotonic support.
\end{definition}

From the definition, we can conclude that comonotonicity is a very strong positive dependency structure. Indeed, if $\underline{x}$ and $\underline{y}$ are elements of the (comonotonic) support of $\underline{X}$, i.e., $\underline{x}$ and $\underline{y}$ are possible outcomes of $\underline{X}$,
then they must be ordered componentwise. This explains why the term comonotonic (common monotonic) is used \cite{dhaene2002concept}.

\begin{theorem} \cite{dhaene2002concept}
A random vector $\underline{X}=(X_1,X_2,\ldots,X_n)$ is comonotonic if and only if one of the following equivalent conditions holds: 
\begin{description}
\item (1)
$\underline{X}$ has a comonotonic support.
\item (2)
For all $\underline{x}=(x_1,x_2,\ldots,x_n)$, we have
\begin{equation}
F_{\underline{X}}(\underline{x}) = \min\{ F_{X_1}(x_1), F_{X_2}(x_2),\ldots, F_{X_n}(x_n) \}.
\end{equation}
\item (3)
For $U\sim Uniform(0,1)$, we have
\begin{equation}
\underline{X} \stackrel{d}{=} (F_{X_1}^{-1}(U),F_{X_2}^{-1}(U),\ldots,F_{X_n}^{-1}(U)).
\end{equation}
\item (4)
There exist a random variable $Z$ and non-decreasing functions $f_i (i=1,2,\ldots,n)$, such that
\begin{equation}
\underline{X} \stackrel{d}{=} (f_{1}(Z),f_{2}(Z),\ldots,f_{n}(Z)).
\end{equation}
\end{description}
\end{theorem}

\subsubsection{Independence}

\begin{theorem} \cite{embrechts2005quantitative,nelsen2006introduction}
Random variables with continuous distributions are independent if and only if their dependence structure is given by 
\begin{equation}
C(u_1,\ldots,u_d) = \prod_{i=1}^{d}u_i.
\end{equation} 
\end{theorem}

\subsubsection{Markovian}

The Markov property is a pure dependence property that can be formulated exclusively in terms of copulas, as a consequence, starting with a Markov process, a multitude of other Markov processes can be constructed by just modifying the marginal distributions \cite{darsow1992copulas,overbeck2015multivariate}. 


\begin{definition} \cite{darsow1992copulas,cherubini2008copula,cherubini2011copula}
Assume two bivariate copulas $A(u,t)$ and $B(t,v)$: the product
operator $\ast$ is defined as
\begin{equation}
A\ast{B}(u,v) \equiv \int_{0}^{1}\frac{\partial{A(u,t)}}{\partial{t}}\frac{\partial{B(t,v)}}{\partial{t}} dt.
\end{equation}
Assume an m-dimensional copula $A$ and an n-dimensional copula $B$: the start operator $\star$ is defined as
\begin{equation}
A\star{B}(u_1,u_2,\ldots,u_{m+n-1}) \equiv \int_{0}^{u_m}\frac{\partial{A(u_1,\ldots,u_{m-1},t)}}{\partial{t}}\frac{\partial{B(t,u_{m+1},\ldots,u_{m+n-1})}}{\partial{t}} dt.
\end{equation}
\end{definition}

\begin{theorem} \cite{darsow1992copulas,cherubini2008copula,cherubini2011copula}
A real valued stochastic process $X_t$ is a Markov process of first order if and only if for all positive integers $n$ and for all $t_1,\ldots,t_n$ satisfying $t_k<t_{k+1}$, $k=1,\ldots,n-1$,
\begin{equation}
C_{t_1,\ldots,t_n}=C_{t_1,t_2}\star C_{t_2,t_3}\star\ldots\star C_{t_{n-1},t_n},
\end{equation}
where $C_{t_1,\ldots,t_n}$ is the copula of $X_{t_1},\ldots,X_{t_n}$ and $C_{t_k,t_{k+1}}$ is the copula of $X_{t_k}$ and $X_{t_{k+1}}$.
\end{theorem}

\nop{
\begin{proposition}
Let $X$ and $Y$ be two real-valued random variables with a dependence
structure represented by the copula function $C_{X,Y}$ and continuous marginal distributions $F_X$ and $F_Y$. Then,
\begin{eqnarray}
C_{X,X+Y}(u,v) &=& \int_{0}^{1}D_{1}C_{X,Y}(w,F_Y(F_{X+Y}^{-1}(v)-F_X^{-1}(w)))dw, \\
F_{X+Y}(t) &=& \int_{0}^{1}D_{1}C_{X,Y}(w,F_Y(t-F_X^{-1}(w)))dw.
\end{eqnarray}
\end{proposition}
}

A strictly stationary Markov process is characterized by assuming $C_{X_{i-1},X_i}\equiv C$, $\forall{i}$ and $F_{X_i}\equiv F$, $\forall{i}$. This kind of process can be constructed for any given $C$ and $F$, with $D_{1}C(u,v)=\frac{\partial{C}}{\partial{u}}(u,v)$, by setting
\begin{equation}
F_{X_i}(t) = \int_{0}^{1}D_{1}C(w,F(t+F^{-1}(w)))dw \equiv G(t),
\end{equation}
and 
\begin{equation}
C_{X_{i-1},X_i}(u,v) = \int_{0}^{u}D_{1}C(w,F(G^{-1}(t)+F^{-1}(w)))dw.
\end{equation}

\begin{definition} \cite{overbeck2015multivariate}
Assume that, for $\mathbf{x}\in[0,1]^k$ the distribution $C(\mathbf{x},.)$ is absolutely continuous with respect to the measure generated by some copula $A$ on $[0, 1]^l$. We denote by $C_{,A}(\mathbf{x}, \mathbf{y})$ (a version of) theRadon-Nikodym derivative,
\begin{equation}
C(\mathbf{x},d\mathbf{y}) = C_{,A}(\mathbf{x},\mathbf{y})A(d\mathbf{y}).
\end{equation}
The subscript ``, $A$'' indicates that we take the derivative with respect to the second set of arguments $(\mathbf{y})$. Accordingly, we define the derivative $C_{B,}(\mathbf{x}, \mathbf{y})$ of $C$ with respect to a $k$-dimensional copula $B$ by
\begin{equation}
C(d\mathbf{x},\mathbf{y}) = C_{B,}(\mathbf{x},\mathbf{y})B(d\mathbf{x}),
\end{equation}
provided that for given $\mathbf{y}$ the measure generated by $C(., \mathbf{y})$ is absolutely continuous with respect to the measure
generated by $B$. $C_{,A}(\mathbf{x}, \mathbf{y})$ and $C_{B,}(\mathbf{x}, \mathbf{y})$ are called derivative of the copula $C(\mathbf{x}, .)$ resp. $C(., \mathbf{y})$ with respect to the copula $A$ resp. $B$.
\end{definition}

\begin{definition} \cite{overbeck2015multivariate}
Let $A$ be a $(k + m)$-dimensional copula, $B$ be a $(m + l)$-dimensional
copula and $C$ be a $m$-dimensional copula such that the derivatives $A_{,C}$ and $B_{C,}$ are well-defined. The operator $\stackrel{C(.)}{\star}$ is defined by
\begin{equation}
(A\stackrel{C(.)}{\star}B)(\mathbf{x},\mathbf{y}) = \int_{0}^{\mathbf{z}} A_{,C}(\mathbf{x},\mathbf{r})\cdot B_{C,}(\mathbf{r},\mathbf{y})C(d\mathbf{r}),
\end{equation}
provided that the integral exists for all $\mathbf{x}$, $\mathbf{y}$, $\mathbf{z}$.
\end{definition}

\begin{theorem} \cite{overbeck2015multivariate}
The $n$-dimensional process $\mathbf{X}$ is a Markov process if and only if for all $t_1< t_2<\ldots < t_p$ the copula $C^{t_1,\ldots,t_p}$ of $(\mathbf{X}_{t_1},\ldots,\mathbf{X}_{t_p})$ satisfies
\begin{equation}
C^{t_1,\ldots,t_p} = C^{t_1,t_2}\stackrel{C^{t_2}(.)}{\star}C^{t_2,t_3}\stackrel{C^{t_3}(.)}{\star}\ldots\stackrel{C^{t_{p-1}}(.)}{\star}C^{t_{p-1},t_p}.
\end{equation}
\end{theorem}

Given a Markov family $C^{st}$, $s, t\in\mathscr{T}$, $s < t$, we can define finite dimensional copulas $C^{t_1,\ldots,t_p}$ and combine these with an arbitrarily specified flow $\mathbf{F}_t(\mathbf{x})=(F_{X_t^1}(x_1),\ldots,F_{X_t^n}(x_n))$ of marginal one-dimensional distributions $F_{X_t^i}$
to obtain finite dimensional distributions
\begin{equation}
\mathbf{P}\left( \mathbf{X}_{t_1}<\mathbf{x}_1,\ldots,\mathbf{X}_{t_p}<\mathbf{x}_p \right) = C^{t_1,\ldots,t_p}\left( \mathbf{F}_{t_1}(\mathbf{x}_1),\ldots,\mathbf{F}_{t_p}(\mathbf{x}_p) \right).
\end{equation}
By applying Kolomogorov's construction theorem for stochastic processes there exists a Markov process $\mathbf{X}$ with the given copulas and marginal distributions.

\begin{remark}
The copula representation of Markov processes of order $k$ has been extended in \cite{ibragimov2005copula,ibragimov2009copula}, and the construction of Markov processes through increment aggregation is elaborated in \cite{cherubini2011copula}. Moreover, the results for the one-dimensional case in \cite{darsow1992copulas} are extended to the general multivariate setting in \cite{overbeck2015multivariate}. In addition, the relationship between Markov property and martingale is investigated in \cite{ibragimov2005copula,cherubini2008copula,cherubini2011copula}.
\end{remark}


\begin{theorem} \cite{cherubini2008copula,cherubini2011copula}
Let $X=\{X_i\}_{i\ge{0}}$ be a Markov process and set $Y_i=X_i-X_{i-1}$. $X$ is a martingale if and only if:
\begin{description}
\item{(1)}
$F_{Y_i}$ has finite mean for every $i$;
\item{(2)}
for $i\ge{1}$, $\int_{0}^{1}F_{Y_i}^{-1}(v)dD_{1}C_{X_{i-1},Y_i}(u,v)=0$, $\forall{u}\in[0,1]$.
\end{description}
\end{theorem}

\begin{proposition} \cite{cherubini2008copula,cherubini2011copula}
Any process for which the distribution of increments is symmetric and the copula between the increments and the levels is symmetric (around the first coordinate) is a martingale.
\end{proposition}

By definition, $(X,Z)$ is a martingale with respect to $\mathscr{F}^{X,Z}$ iff $\forall{i}\ge{0}$
\begin{eqnarray}
\mathbb{E}[X_{i+1}-X_i|X_i,Z_i] &=& 0, \\
\mathbb{E}[Z_{i+1}-Z_i|X_i,Z_i] &=& 0.
\end{eqnarray}

Let $\Delta X_i=X_{i+1}-X_i$, $\Delta Z_i=Z_{i+1}-Z_i$ and $A_{i,i+1}(u,v,w,\lambda)$ be the copula function of the random vector $(X_i,Z_i,\Delta{X_i},\Delta{Z_i})$ with $F_{X_i}$, $F_{Z_i}$, $F_{\Delta{X_i}}$, $F_{\Delta{Z_i}}$ the corresponding marginal cdf. Set $a_{i,i+1}(u,v,w,1)$ the density of the copula $A_{i,i+1}(u,v,w,1)$ and $a_{i,i+1}(u,v,1,w)$ the density of the copula $A_{i,i+1}(u,v,1,w)$.

\begin{theorem} \cite{cherubini2011copula}
The bivariate Markov process $(X,Z)$ is a martingale with respect to the filtration $\mathscr{F}^{X,Z}$ iff:
\begin{description}
\item{(1)}
$F_{\Delta{X_i}}$ and $F_{\Delta{Z_i}}$ have finite mean for every $i$;
\item{(2)}
for every $i$, $\forall u,v\in[0,1]$,
\begin{eqnarray}
\int_{0}^{1}F_{\Delta{X_i}}^{-1}(w)a_{i,i+1}(u,v,w,1)dw &=& 0, \\
\int_{0}^{1}F_{\Delta{Z_i}}^{-1}(w)a_{i,i+1}(u,v,1,w)dw &=& 0.
\end{eqnarray}
\end{description}
\end{theorem}

\section{Analysis of Cumulative Capacity}\label{analysiscc}

\subsection{General Results}

\subsubsection{Exact Expression}

The CDF of the cumulative capacity can be derived from the joint distribution of the channel gain magnitude, i.e.,
\begin{equation}
F_{S(s,t)}(x) = \int_{S(s,t) = \sum_{i=s+1}^{t} \log_2(1+\gamma|h_i|^2) \le{x}} f_{\textbf{H}}(h_{s+1},h_{s+2},\ldots,h_{t}) d\textbf{H}.
\end{equation}
For example, the single-integral form PDF and CDF of the multivariate generalized Rician
distribution are expressed as \cite{beaulieu2011novel}
\begin{eqnarray}
f_{\textbf{H}}(h_1,h_2,\ldots,h_N) &=& \int_{t=0}^{\infty}\frac{t^{\frac{m-1}{2}}}{S^{m-1}}\exp(-(t+S^2))I_{m-1}(2S\sqrt{t})\prod_{k=1}^{N}\frac{1}{(\lambda_k^2\sigma_k^2{t})^{\frac{m-1}{2}}} \nonumber\\
&& \times \frac{1}{\Omega_k^2}{h_{k}^{m}}\exp\left( -\frac{h_k^2+\lambda_k^2\sigma_k^2{t}}{2\Omega_k^2} \right) I_{m-1}\left( h_k\frac{\sqrt{\sigma_k^2\lambda_k^2{t}}}{\Omega_k^2} \right)dt, \\
F_{\textbf{H}}(h_1,h_2,\ldots,h_N) &=& \int_{t=0}^{\infty}\frac{t^{\frac{m-1}{2}}}{S^{m-1}}\exp(-(t+S^2))I_{m-1}(2S\sqrt{t}) \prod_{k=1}^{N}\left[ 1- Q_{m}\left( \frac{\sqrt{t}\sqrt{\sigma_k^2\lambda_k^2}}{\Omega_k},\frac{h_k}{\Omega_k} \right) \right]dt,
\end{eqnarray}
where $\Omega_k^2=\sigma_k^2\left(\frac{1-\lambda_k^2}{2}\right)$ and $S^2=\sum_{l=1}^{m}(m_{1l}^{2}+m_{2l}^{2})$.

According to the transform of random vector that is elaborated in Theorem \ref{vectortrans}, the CDF can also be expressed as 
\begin{equation}
F_{S(s,t)}(x) = \int_{S(s,t)=\sum_{i=s+1}^{t}c_i \le{x}} f_{\textbf{C}}(c_{s+1},c_{s+2},\ldots,c_t) d\textbf{C},
\end{equation}
where $f_\textbf{C}(\textbf{y}) = f_{\textbf{H}}(C^{-1}(\textbf{y}))|J_0(\textbf{y})|$ and $C(\textbf{y})=\log_2(1+\gamma|\textbf{H}(\textbf{y})|^2)$.

\begin{theorem}\label{vectortrans} \cite{fessler1998transformations}
Let $g: \mathbb{R}^n\rightarrow\mathbb{R}^n$ be one-to-one and assume that $h=g^{-1}$ is continuous. Assume that on an open set $\mathcal{V}\subseteq\mathbb{R}^n$ $h$ is continuously differentiable with Jacobian $J(y)$. Define $J_0: \mathbb{R}^n\rightarrow\mathbb{R}^n$ by
\begin{equation}
J_0(y) = \begin{cases}
J(y), & y\in\mathcal{V} \\
0, & y\in\mathcal{V}^c,
\end{cases}
\end{equation}
where $\mathcal{V}^c$ is the set complement (in $\mathbb{R}^n$) of $\mathcal{V}$.
Suppose random vector $X$ has pdf $f_X(x)$ (with respect to Lebesgue measure) with nonzero mass in $h(\mathcal{V}^c)$, i.e., $P\{ X\in h(\mathcal{V}^c) \}=\int_{\mathcal{V}^c} f_X(x)dx =0$. Then the pdf of $Y=g(X)$ is given by 
\begin{equation}
f_Y(y) = f_X(g^{-1}(y))|J_0(y)| = \begin{cases}
f_X(g^{-1}(y))|J(y)|, & y\in\mathcal{V} \\
0, & y\in\mathcal{V}^c.
\end{cases}
\end{equation}
\end{theorem}

In addition, the exact formula can also be expressed with copula. 
For a $d$-dimensional random vector $\mathbf{X}$, denote $\mu_C$ the measure on $[0,1]^d$ implied by the copula $C$, that is $\mu_C(B):=\mathbb{P}[\mathbf{U}\in{B}]$ for any Borel measurable $B\subseteq[0,1]^d$ and $\mathbf{U}\sim C$. Define $Z=\sum_{i=1}^{d}\omega_{i}x_i$, let $A_Z$ be the convex set $A_Z=\{ \mathbf{x}\in\mathbb{R}^d: \sum_{i=1}^{d}\omega_{i}x_i\le{z} \}$, and $A_Z^{\ast}$ be the linear boundary connected to $A_Z$, i.e., $A_Z^\ast=\{ \mathbf{x}\in\mathbb{R}^d: \sum_{i=1}^{d}\omega_{i}x_i={z} \}$.

\begin{proposition} \cite{gijbels2014distribution}
\begin{equation}
F_Z(z) = \mu_C(B_Z) = \int_{B_Z} c d\lambda,
\end{equation}
where $c$ is the density of $C$, $B_Z=\varphi(A_Z)\subseteq[0,1]^d$, and $\varphi:\mathbb{R}^d\rightarrow[0,1]^d,(x_1,\ldots,x_d)\mapsto(F_1(x_1),\ldots,F_d(x_d))$.
\end{proposition}

\begin{proposition} \cite{gijbels2014distribution}
Assume that $f_i>0$ on $\mathbb{R}$ for $i=1,\ldots,d$, then
\begin{description}
\item{(1)}
the set $B_z^\ast=\varphi(A_z^\ast)$ is given by $B_z^\ast=\{ \mathbf{u}\in[0,1]^d: (u_1,\ldots,u_{d-1})\in(0,1)^{d-1}, u_d=\tau_z(u_1,\ldots,u_{d-1}) \}$, where $\tau_z:(0,1)^{d-1}\rightarrow(0,1),\mathbf{u}\mapsto F_d(\frac{z}{\omega_d}-\sum_{i=1}^{d-1}\frac{\omega_i}{\omega_d}F_i^{-1}(u_i))$;
\item{(2)}
the set $B_z$ is given as $B_z=\{ \mathbf{u}\in[0,1]^d: (u_1,\ldots,u_{d-1})\in(0,1)^{d-1}, 0<u_d\le\tau_z(u_1,\ldots,u_{d-1}) \}$;
\item{(3)}
for $z_1<z_2$, $B_{z_1}\subsetneq B_{z_2}$;
\item{(4)}
$\tau_z(\mathbf{u})$ is a strictly decreasing function in each component of $\mathbf{u}$;
\item{(5)}
$B_z$ is path-connected and $B_z^\epsilon:=B_z\cap[\epsilon,1-\epsilon]^d$ is start-shaped for $\epsilon>0$ with center $(\epsilon,\ldots,\epsilon)$.
\end{description}
\end{proposition}

\begin{remark}
Algorithms for numerical calculation of $F_Z(z)$ are elaborated in \cite{gijbels2014distribution,arbenz2012gaep,arbenz2011aep}.
\end{remark}

\subsubsection{Bounds}

In the following, instead of trying to find accurate representation of $F_{S(s,t)}$, we turn to investigate bounds on it. 
With Theorem \ref{copula}, it is easily verified that the CDF of the channel cumulative capacity satisfies the following inequalities: 
\begin{equation}
F_{S(s,t)}^{l}(r)\le F_{S(s,t)}(r)\le F_{S(s,t)}^{u}(r), \label{dcdf}
\end{equation}
where 
\begin{eqnarray}
F_{S(s,t)}^{u}(r) &\equiv& \inf_{\sum\limits_{i=s+1}\limits^{t}r_{i}=r}\left[
\sum\limits_{i=s+1}\limits^{t}{F_{C(i)}(r_{i})}\right]_{1},
\label{dtccdf}\\
F_{S(s,t)}^{l}(r) &\equiv& \sup_{\sum\limits_{i=s+1}\limits^{t}r_{i}=r}\left[
\sum\limits_{i=s+1}\limits^{t}{F_{C(i)}(r_{i})} -(t-s-1) \right]^{+}. \label{dtccdfl}
\end{eqnarray}
The improved results with dual bounds can also been obtained following Theorem \ref{dualbasic} to Theorem \ref{dualsharp}.

\subsection{Special Cases}

\subsubsection{Comonotonicity}

The copula of a distribution function contains all the dependence information, specifically for a comonotonic dependence structure, i.e., comonotonic random variables are increasing functions of a common random variable \cite{embrechts2003using}, the joint distribution function of the instantaneous capacity is expressed as \cite{dhaene2002concept,embrechts2006bounds}
\begin{equation}
F(C(s+1),\ldots,C(t)) = \min\left\{ F(C(s+1)),\ldots,F(C(t)) \right\},
\end{equation}
and the CDF of the cumulative capacity is expressed as \cite{dhaene2002concept,embrechts2006bounds}
\begin{equation}
m_{t-s}(x) = \int_{\sum_{i=s+1}^{t}C(i)<x} d{ \min\left\{ F(C(s+1)),\ldots,F(C(t)) \right\} }.
\end{equation}

In the special case that all marginal distribution functions are identical $F_{C(i)}\sim F_{C}$, comonotonicity of $C(i)$ is equivalent to saying that $C(s+1)=C(s+2), \ldots, =C(t)$ holds almost surely \cite{dhaene2002concept}, in other words, the capacity depends only on the initial state and keeps constant afterward, i.e.,
\begin{equation}
F_{S(s,t)}(x) = F_{C}\left( \frac{x}{t-s} \right).
\end{equation}

\subsubsection{Independence}

If $C(i)$ and $C(j)$, $i\neq j$, are independent, $f_{S(s,t)} = f_{C(s+1)} \ast \ldots \ast f_{C(t)}$, where $\ast$ denotes the convolution operation.
Hence, 
\begin{equation}
F_{S(s,t)}(x) = \int_{-\infty}^{x} f_{S(s,t)}(y) dy.
\end{equation}
In addition, when the length of the period, $t-s$, is large, according to the central limit theorem, $F_{S(s,t)}(x)$ approaches a normal distribution with mean $E[S(s,t)]$ and variance $\sigma^{2}[S(s,t)]$(under certain general conditions) \cite{papoulis2002probability}, i.e.,
\begin{eqnarray}\label{tccdf}
F_{S(s,t)}(x) \approx G\left(\frac{x-E[S(s,t)]}{\sigma^{2}[S(s,t)]}\right), 
\end{eqnarray}
where $E[S(s,t)]= \sum\limits_{i=s+1}\limits^{t}E[C(i)]$, 
$\sigma^{2}[S(s,t)]= \sum\limits_{i=s+1}\limits^{t}\sigma^{2}[C(i)]$, and $G(x) \equiv \int_{-\infty}^{x}\frac{1}{\sqrt{2\pi}}e^{-y^2/2}dy$. 

In the special case that all marginal distribution functions are identical $F_{C(i)}\sim F_{C}$, the cumulant generating function and the likelihood ratio are expressed as \cite{asmussen2003applied}
\begin{eqnarray}
\kappa(\theta) &=& \log\mathbb{E}e^{\theta{C(i)}} = \log\int e^{\theta{x}} F(dx), \\
L_t &=& e^{\theta{S_t}-t\kappa(\theta)},
\end{eqnarray}
where $L_t$ is a mean-one martingale.
According to the Markov inequality
\begin{equation}
P\{ L_t\ge{\mu} \} \le\frac{1}{\mu}\mathbb{E}[L_t]=\frac{1}{\mu},
\end{equation}
i.e., for $\theta<0$
\begin{equation}
P\{ S_t\ge x \} \le e^{\theta{x}-t\kappa(\theta)}.
\end{equation}

\subsubsection{Markovian}

The likelihood ratio martingale can be used to provide exponential upper bounds \cite{gallager2013stochastic}. Define
\begin{equation}
\underline{L}_n = \frac{\min_n(h^{(\theta)}(J_n))}{h^{(\theta)}(J_0)}e^{-\theta{S_n}+n\kappa(\theta)}.
\end{equation}
Then $\underline{L}_n\le L_n$, so $\mathbb{E}[\underline{L}_n]\le{1}$. For any $\mu>0$, the Markov inequality and Doob's submartingale inequality can be applied to $\underline{L}_n$ and $L_n$ to get
\begin{eqnarray}
P\{ \underline{L}_n\ge{\mu} \} &\le& \frac{1}{\mu}\mathbb{E}[\underline{L}_n]\le \frac{1}{\mu}, \\
P\left\{ \sup_{n\ge{1}}{L_n} \ge \mu \right\} &\le& \frac{1}{\mu}\mathbb{E}[{L}_n]\le \frac{1}{\mu}.
\end{eqnarray}
For any $\alpha$ and $\theta<0$, we can choose $\mu=e^{n\kappa(\theta)-\theta\alpha}\min_{n}(h^{(\theta)}(J_n))/h^{(\theta)}(J_0)$
\begin{eqnarray}
P\{ S_n\ge\alpha \} \le \frac{h^{(\theta)}(J_0)}{\min_{n}(h^{(\theta)}(J_n))}e^{-n\kappa(\theta)+\theta\alpha}.
\end{eqnarray}

\section{Analysis of Maximum and Minimum Cumulative Capacity}\label{analysismcc}

\subsection{General Results}

With assumption of non-Granger causality on the dependence structure, a lower bound for the CDF of the maximum cumulative capacity and an upper bound for the CDF of the minimum cumulative capacity are obtained
\begin{eqnarray}
\mathbb{P}\left( \sup_{0\le{i}\le{t}}\overline{S}(i)\le x \right) &=& \mathbb{P}\left( S(1)\le x,S(2)\le x,\ldots, S(t)\le x \right) \\
&\ge& \mathbb{P}\left( \max C(1)\le x,\max_{1\le i\le 2} C(i)\le\frac{x}{2},\ldots, \max_{1\le i\le t} C(i)\le\frac{x}{t}\right) \\
&=& {C}\left({F}_{M_1}(x),{F}_{M_2}\left(\frac{x}{2}\right),\ldots,{F}_{M_t}\left(\frac{x}{t}\right)\right) \\
&=& {C}\left( {F}(x),{F}\left(\frac{x}{2},\frac{x}{2}\right),\ldots,{F}\left(\frac{x}{t},\frac{x}{t},\ldots,\frac{x}{t} \right) \right),
\end{eqnarray}
\begin{eqnarray}
\mathbb{P}\left( \inf_{0\le{i}\le{t}}\underline{S}(i)\le x \right) &=& 1- \mathbb{P}\left( S(1)> x,S(2)> x,\ldots, S(t)> x \right) \\
&\le& 1- \mathbb{P}\left( \min C(1)> x,\min_{1\le i\le 2} C(i)>\frac{x}{2},\ldots, \min_{1\le i\le t} C(i)>\frac{x}{t}\right) \\
&=& 1 - \overline{C}\left(\overline{F}_{m_1}(x),\overline{F}_{m_2}\left(\frac{x}{2}\right),\ldots,\overline{F}_{m_t}\left(\frac{x}{t}\right)\right) \\
&=& 1 - \overline{C}\left( \overline{F}(x),\overline{F}\left(\frac{x}{2},\frac{x}{2}\right),\ldots,\overline{F}\left(\frac{x}{t},\frac{x}{t},\ldots,\frac{x}{t} \right) \right),
\end{eqnarray}
where $F(x_1,x_2,\ldots,x_t)=C(F_{C(1)}(x_1),F_{C(2)}(x_2),\ldots,F_{C(t)}(x_t))$.

\subsection{Special Cases}

\subsubsection{Independence}

In the special case that all marginal distribution functions are identical $F_{C(i)}\sim F_{C}$, the cumulant generating function and the likelihood ratio are expressed as \cite{asmussen2003applied}
\begin{eqnarray}
\kappa(\theta) &=& \log\mathbb{E}e^{\theta{C(i)}} = \log\int e^{\theta{x}} F(dx), \\
L_t &=& e^{\theta{S_t}-t\kappa(\theta)},
\end{eqnarray}
where $L_t$ is a mean-one martingale.
Let the Lundberg equation $\kappa(\theta)=0$ and assume the existence of a solution $\theta>0$, then \cite{asmussen2003applied}
\begin{equation}
P\left\{ \sup_{t\ge{0}}S_t \ge{x} \right\} \le e^{-\theta{x}},
\end{equation}
for all $x\ge{0}$.

\subsubsection{Markovian}

Let $\tau(u)=\inf\{ t>0: S_t>u \}$, $I(u)=J_{\tau(u)}$, $\xi(u)=S_{\tau(u)} - u$, $M=\sup_{t\ge{0}}S_t$. Let the Lundberg equation $\kappa(\theta)=0$ and assume the existence of a solution $\theta>0$.
Then \cite{asmussen2003applied,asmussen2010ruin}
\begin{eqnarray}
\mathbb{P}_i(M>u) &=& \mathbb{P}_{i}(\tau(u)<\infty) = \mathbb{E}_{i,\theta}\left[ \frac{h^{(\theta)}_{J_0}}{h^{(\theta)}_{J_{\theta(u)}}}e^{-\theta S_{\tau(u)}};\tau(u)<\infty \right] \\
&=& e^{-\theta{u}}\mathbb{E}_{i,\theta} \left[ \frac{h^{(\theta)}_{i}}{h^{(\theta)}_{I(u)}}e^{-\theta{\xi(u)}} \right], \\
\mathbb{P}(M>u) &=& \sum_{i}\pi_i\mathbb{P}_i.
\end{eqnarray}
According to Lundberg's inequility \cite{asmussen2010ruin}
\begin{eqnarray}
\mathbb{P}_i(M>u) &\le& \frac{h_i^{(\theta)}}{\min_{j\in{E}}h_j^{(\theta)}}e^{-\theta{u}},\\
\mathbb{P}(M>u) &\le& \sum_{i}\pi_i\mathbb{P}_i.
\end{eqnarray}

The constant in front of $e^{-\theta{u}}$ in Lundberg’s inequality can be improved and a supplementary lower bound can also be obtained with the following theorem.

\begin{theorem} \cite{asmussen2010ruin}
Let 
\begin{eqnarray}
C_{-} &=& \min_{j\in{E}}\frac{1}{h_j^{(\theta)}}\cdot\inf_{x\ge{0}}\frac{\overline{B}_j(x)}{\int_{x}^{\infty}e^{\theta(y-x)}B_j(dy)},\\
C_{+} &=& \max_{j\in{E}}\frac{1}{h_j^{(\theta)}}\cdot\sup_{x\ge{0}}\frac{\overline{B}_j(x)}{\int_{x}^{\infty}e^{\theta(y-x)}B_j(dy)},
\end{eqnarray}
where $B_j$ is the distribution of the increment.
Then for all $j\in{E}$ and all $u\ge{0}$,
\begin{equation}
C_{-}h_i^{(\theta)}{e^{-\theta{u}}}\le \mathbb{P}_i(M>u) \le C_{+}h_i^{(\theta)}{e^{-\theta{u}}}.
\end{equation}
\end{theorem}

\begin{remark}
The above results in this section have been so far essentially for the forward-looking maximum / minimum cumulative capacity. For their backward-looking counterparts, similar analysis may apply, and so may for the the general definitions of maximum and minimum cumulative capacity. 
\end{remark}

\section{Other Characteristics of Cumulative Capacity}\label{othercht}

\subsection{Moment Generating Function}

The moment generating function (MGF) of $S(s,t)=\sum\limits_{i=s+1}\limits^{t}C(i)$, denoted by $M_{S(s,t)}(\theta)$, is \cite{jiang2008stochastic,fidler2006end}
\begin{IEEEeqnarray}{rCl}
M_{S(s,t)}(\theta)&\equiv&{E\left[e^{\theta{S(s,t)}}\right]} 
= \int_{-\infty}^{\infty} {e^{\theta{r}}}d{F_{S(s,t)}(r)},
\end{IEEEeqnarray}
where $\theta$ is a real variable. We denote $\overline{M}_{S(s,t)}(\theta)=M_{S(s,t)}(-\theta)={E\left[e^{-\theta{S(s,t)}}\right]}$.

With $F_{S(s,t)}(x)$ derived in the previous section, $M_{S(s,t)}(\theta)$ is readily obtained. Specifically, for the independent case, we have 
\begin{eqnarray}
\overline{M}_{S(s,t)}^{i}(\theta) &=& \Pi_{i=s+1}^{t}\overline{M}_{C(i)}(\theta) \label{mgf-1}\\
 &\approx& \int_{-\infty}^{\infty} {e^{-\theta{r}}} d G\left(\frac{r-E[S(s,t)]}{\sigma^{2}[S(s,t)]}\right), \label{imgf}
\end{eqnarray}
where for the independent and identically distributed (i.i.d.) case, (\ref{mgf-1}) becomes 
\begin{equation}
\overline{M}_{S(s,t)}^{ii}(\theta) = (\overline{M}_{C(i)}(\theta) )^{t-s}. \label{iimgf}
\end{equation}
For the more general case,
\begin{eqnarray}
\overline{M}_{S(s,t)}^{dl}(\theta) \le \overline{M}_{S(s,t)}^{d}(\theta) \le \overline{M}_{S(s,t)}^{du}(\theta), \label{dmgf}
\end{eqnarray}
with
\begin{eqnarray}
\overline{M}_{S(s,t)}^{dl}(\theta) &=& \int_{-\infty}^{\infty} {e^{-\theta{r}}} d F_{S(s,t)}^{l}(r), \\
\overline{M}_{S(s,t)}^{du}(\theta) &=& \int_{-\infty}^{\infty} {e^{-\theta{r}}} d F_{S(s,t)}^{u}(r).
\end{eqnarray}

\subsubsection{Rayleigh channel as an example}
For the i.i.d. case, 
\begin{IEEEeqnarray}{rCl}
\overline{M}_{S(s,t)}^{ii}(\theta) = \left( \int_{0}^{\infty} {e^{-\theta{r}}} d \left\{ 1-e^{-(2^r-1)/\gamma} \right\} \right)^{t-s} 
= \left( \frac{\ln{2}}{\gamma} \int_{0}^{\infty} {e^{-\theta{r}}} 2^{r}e^{-(2^r-1)/\gamma} d r \right)^{t-s}.
\end{IEEEeqnarray}

For the general (possibly dependent) case, the upper bound and the lower bound of the MGF $\overline{M}^{d}_{S(s,t)}(\theta)$ are expressed as
\begin{IEEEeqnarray}{rCl}
{\overline{M}^{du}_{S(s,t)}(\theta)} 
&=& \int_{0}^{\infty} {e^{-\theta{r}}} d \left\{ \left[ (t-s)\left( 1-e^{-\left(2^{\frac{r}{t-s}}-1\right)\Big{/}\gamma} \right) \right]_{1} \right\} \IEEEeqnarraynumspace\\
&=& (t-s)\left(1-\exp\left(\frac{1-2^{\Omega_{u}/(t-s)}}{\gamma}-\theta{\Omega_{u}}\right)\right) 
-\theta(t-s) \int_{0}^{\Omega_{u}} \exp\left(  \frac{1-2^{r/(t-s)}}{\gamma}-\theta{r}\right) d r, \IEEEeqnarraynumspace\\
{\overline{M}^{dl}_{S(s,t)}(\theta)}
&=& \int_{0}^{\infty} {e^{-\theta{r}}} d \left\{ \left[ 1-(t-s)e^{-\left(2^{\frac{r}{t-s}}-1\right)\Big{/}\gamma}  \right]^{+} \right\} \IEEEeqnarraynumspace\\
&=& (t-s)\exp\left(\frac{1-2^{\Omega_{l}/(t-s)}}{\gamma}-\theta{\Omega_{l}}\right) 
-\theta(t-s) \int_{\Omega_{l}}^{\infty} \exp\left(  \frac{1-2^{r/(t-s)}}{\gamma}-\theta{r}\right) d r,
\end{IEEEeqnarray}
where $\Omega_{u}=\tau\log_{2}\left(  1-\gamma\log(1-1/\tau)\right)$ for $\tau>1$, $\Omega_{l}=\tau\log_{2}(1-\gamma\log(1/\tau))$ for $\tau\ge 1$, and $\tau=t-s$.

\subsubsection{Application to effective capacity}

Suppose the service process has stationary increments, i.e., $S(s,s+t)=_{st}S(t)$ for all $s,t\le 0$, the effective capacity of $S$ is defined as \cite{jiang2008stochastic,wu2003effective}
\begin{eqnarray}
\mathring{r}^{(c)}(\theta) \equiv -{{\lim\sup}_{t\rightarrow\infty}\frac{1}{\theta{t}}\sup_{s\geq{0}}\log{E\left[e^{-\theta(S(s,s+t))}\right]}},
\end{eqnarray}
where $\theta$ is a real variable.

The effective capacity calculated through the instantaneous capacity follows directly from (\ref{imgf}) and (\ref{dmgf}), i,e.,
\begin{IEEEeqnarray}{rCl}
\mathring{r}^{(c)}_{i}(\theta) &=& -{\lim\sup}_{t\rightarrow\infty}\frac{1}{\theta{t}}\sup_{s\geq{0}}\log{\overline{M}_{S(s,s+t)}^{i}(\theta)},\IEEEeqnarraynumspace\\
\mathring{r}^{(c)}_{d}(\theta) &=& -{\lim\sup}_{t\rightarrow\infty}\frac{1}{\theta{t}}\sup_{s\geq{0}}\log{\overline{M}_{S(s,s+t)}^{d}(\theta)},\IEEEeqnarraynumspace
\end{IEEEeqnarray}
for the independent and dependent cases respectively.

For the Rayleigh fading channel, $\frac{1}{\theta{\tau}}\log{\overline{M}_{S(\tau)}^{ii}(\theta)}$ is constant with $\tau$, $\frac{1}{\theta{\tau}}\log{\overline{M}_{S(\tau)}^{du}(\theta)}$ increases with $\tau$, while $\frac{1}{\theta{\tau}}\log{\overline{M}_{S(\tau)}^{dl}(\theta)}$ decreases with $\tau$. The effective capacity for the i.i.d. case and dependent case are respectively,
\begin{IEEEeqnarray}{rCl}
\mathring{r}^{(c)}_{ii}(\theta) &=& -\frac{1}{\theta}\log{ \left( \frac{\ln{2}}{\gamma} \int_{0}^{\infty} {e^{-\theta{r}}} 2^{r}e^{-(2^r-1)/\gamma} d r \right) }, \IEEEeqnarraynumspace\\
\mathring{r}^{(c)}_{dl}(\theta) &=& -\lim_{\tau\rightarrow\infty}\frac{1}{\theta{\tau}}\log \Big(  \tau(1-\exp(({1-2^{\Omega_{u}/{\tau}}})/{\gamma}
-\theta{\Omega_{u}}))-\theta\tau \int_{0}^{\Omega_{u}} \exp\left(  ({1-2^{r/\tau}})/{\gamma}-\theta{r}\right) d r  \Big) ,\IEEEeqnarraynumspace\\
\mathring{r}^{(c)}_{du}(\theta) &=& -\frac{1}{\theta}\log \Big(  \tau\exp\left(({1-2^{\Omega_{l}/\tau}})/{\gamma}-\theta{\Omega_{l}}\right)  
-\theta\tau \int_{\Omega_{l}}^{\infty} \exp\left( ({1-2^{r/\tau}}){\gamma}-\theta{r}\right) d r  \Big).\IEEEeqnarraynumspace
\end{IEEEeqnarray}

\nop{
\begin{IEEEeqnarray}{rCl}
\IEEEeqnarraymulticol{3}{l}
{\mathring{r}^{(c)}_{ii}(\theta) = -\frac{1}{\theta}\log{ \left( \frac{\ln{2}}{\gamma} \int_{0}^{\infty} {e^{-\theta{r}}} 2^{r}e^{-(2^r-1)/\gamma} d r \right) },} \IEEEeqnarraynumspace\\
\IEEEeqnarraymulticol{3}{l}
{\mathring{r}^{(c)}_{dl}(\theta) = -\lim_{\tau\rightarrow\infty}\frac{1}{\theta{\tau}}\log \Bigg(  \tau\Bigg(1-\exp\Bigg(\frac{1-2^{\Omega_{u}/{\tau}}}{\gamma}}\nonumber\\\qquad\qquad
-\theta{\Omega_{u}}\Bigg)\Bigg)-\theta\tau \int\limits_{0}^{\Omega_{u}} \exp\left(  \frac{1-2^{r/\tau}}{\gamma}-\theta{r}\right) d r  \Bigg) ,\IEEEeqnarraynumspace\\
\IEEEeqnarraymulticol{3}{l}
{\mathring{r}^{(c)}_{du}(\theta) = -\frac{1}{\theta}\log \Bigg(  \tau\exp\left(\frac{1-2^{\Omega_{l}/\tau}}{\gamma}-\theta{\Omega_{l}}\right) } \nonumber\\
-\theta\tau \int\limits_{\Omega_{l}}^{\infty} \exp\left(  \frac{1-2^{r/\tau}}{\gamma}-\theta{r}\right) d r  \Bigg).\IEEEeqnarraynumspace
\end{IEEEeqnarray}
}

\subsection{Mellin Transform} 

If the channel capacity is expressed with the natural logarithm, i.e.,
\begin{equation}
C(t)=\log(g(\gamma_t))~\text{nats/s/Hz},
\end{equation}
in order to circumvent the calculation of logarithm, the cumulative capacity may be transformed to a new domain, e.g., the MGF domain, with the exponential,
\begin{equation}
\mathcal{S}(s,t)=e^{S(s,t)}=\prod\limits_{i=s+1}^{t}g(\gamma_i).
\end{equation}
Then the MGF property analysis applies.

Analog to the MGF for the cumulative capacity, Mellin transform (MT) has also been proposed \cite{al2013min}, i.e.,
\begin{IEEEeqnarray}{rCl}
\IEEEeqnarraymulticol{3}{l}
{\mathcal{M}_{\mathcal{S}(s,t)}(\vartheta) \equiv E[{\mathcal{S}^{\vartheta-1}(s,t)}]} \label{rmt}\\
= \int_{-\infty}^{\infty} { r^{\vartheta-1} } d F_{\mathcal{S}(s,t)}(r) = \int_{-\infty}^{\infty} { r^{\vartheta-1} } d F_{S(s,t)}(r^\ast),
\end{IEEEeqnarray}
where $r^{\ast}=\log_{2}{e}\log{r}$, for any complex variable $\vartheta$ for which the right hand side of (\ref{rmt}) exits.

According to (\ref{tccdf}) and (\ref{dcdf}), 
the MT of the cumulative capacity for the independent and dependent case can be derived,
\begin{IEEEeqnarray}{rCl}
\IEEEeqnarraymulticol{3}{l}
{\mathcal{M}_{\mathcal{S}(s,t)}^{i}(\vartheta) \approx \int_{-\infty}^{\infty} { r^{\vartheta-1} } d G\left(\frac{r^\ast-E[S(s,t)]}{\sigma^{2}[S(s,t)]}\right),} \label{mellini}\IEEEeqnarraynumspace\\
\IEEEeqnarraymulticol{3}{l}
{\mathcal{M}_{\mathcal{S}(s,t)}^{dl}(\vartheta)=\int_{-\infty}^{\infty} {  r^{\vartheta-1}} d F_{S(s,t)}^{l}(r^\ast)\le \mathcal{M}_{\mathcal{S}(s,t)}^{d}(\vartheta)} \label{mellindl}\IEEEeqnarraynumspace\\\qquad\qquad\quad
 \leq \int_{-\infty}^{\infty} { r^{\vartheta-1} } d F_{S(s,t)}^{u}(r^\ast)= \mathcal{M}_{\mathcal{S}(s,t)}^{du}(\vartheta), \label{mellindu}
\end{IEEEeqnarray}
where (\ref{mellindl}) and (\ref{mellindu}) hold when $\vartheta-1< 0$.

\subsubsection{Rayleigh channel as an example}

For the i.i.d. case \cite{al2013min}, 
\begin{equation}
\mathcal{M}^{ii}_{\mathcal{S}(s,t)}(\vartheta)=\left( e^{1/\gamma}\gamma^{\vartheta-1} \int_{ \gamma^{-1} }^{\infty} r^{\vartheta-1}e^{-r} dr \right)^{t-s}.
\end{equation}

For the dependent case, 
\begin{IEEEeqnarray}{rCl}
{\mathcal{M}_{\mathcal{S}(s,t)}^{du}(\vartheta)}  
&=& \int_{0}^{\infty} {r^{\vartheta-1}} d \left\{ \left[ (t-s)\left( 1-e^{-\left(2^{\frac{r^\ast}{t-s}}-1\right)\Big{/}\gamma} \right) \right]_{1} \right\} \IEEEeqnarraynumspace\\
&=& \frac{\log{2}\log_2{e}}{\gamma}\times \int_{1}^{e^{\Omega_u/\log_2{e}}} { r^{\vartheta-2}\times 2^{\frac{r^\ast}{t-s}}\times e^{\frac{1-2^{\frac{r^\ast}{t-s}}}{\gamma}} } dr,
\end{IEEEeqnarray}
\begin{IEEEeqnarray}{rCl}
{\mathcal{M}_{\mathcal{S}(s,t)}^{dl}(\vartheta)} 
&=& \int_{0}^{\infty} {r^{\vartheta-1}} d \left\{ \left[ 1-(t-s)e^{-\left(2^{\frac{r^\ast}{t-s}}-1\right)\Big{/}\gamma}  \right]^{+} \right\} \label{cctdl_alt}\IEEEeqnarraynumspace\\
&\geq& \int_{0}^{\infty} {r\times e^{(\vartheta-2)r}} d \left\{ \left[ 1-(t-s)e^{-\left(2^{\frac{r^\ast}{t-s}}-1\right)\Big{/}\gamma}  \right]^{+} \right\} \IEEEeqnarraynumspace\\
&=& \frac{\log{2}\log_2{e}}{\gamma}\times \int_{ e^{\Omega_l/\log_2{e}} }^{\infty} { e^{(\vartheta-2)r}\times 2^{\frac{r^\ast}{t-s}}\times e^{\frac{1-2^{\frac{r^\ast}{t-s}}}{\gamma}} } dr,
\end{IEEEeqnarray}
where $\vartheta-1<0$, $\Omega_{u}=\tau\log_{2}\left(  1-\gamma\log(1-1/\tau)\right)$ for $\tau>1$, $\Omega_{l}=\tau\log_{2}(1-\gamma\log(1/\tau))$ for $\tau\ge 1$, and $\tau=t-s$.

\subsection{Stochastic Strict Service Curve}

A system is said to be a stochastic strict server providing stochastic strict service curve (SSSC) $\beta$ with
bounding function $g$, if during any period $(s, t]$ the amount of service
$S(s, t)$ provided by the system satisfies \cite{jiang2008stochastic}, 
\begin{eqnarray}
P\left\{S(s,t)<\beta(t-s)-x\right\}\leq g(x),
\end{eqnarray}
for all $x\ge 0$.

\subsubsection{Independent Case}

If $C(i)$ and $C(j)$, $i\neq j$, are independent,
applying (\ref{tccdf}), the expression of SSSC is obtained
\begin{equation}\label{sissc}
P\left\{S(s,t)\leq{\beta(t-s)-x}\right\} 
\approx G\left(\frac{(\beta(t-s)-x)-E[S(s,t)]}{\sigma^{2}[S(s,t)]}\right)=g^{i}_{s,t}(x).
\end{equation}

If $S(s,t)$ also has stationary increments, i.e., $C(i)=_{st}C(j)$, the SSSC can be expressed as \cite{jiang2010note}
\begin{eqnarray}
\beta^{i}(t-s) = \frac{1}{-\theta}\log{\overline{M}_{S(s,t)}^{i}(\theta)} 
= \frac{t-s}{-\theta}\log{\overline{M}_{C}(\theta) }, \label{istationary}
\end{eqnarray}
with bounding function $g(x)=e^{-\theta x}$.

\subsubsection{Dependent Case}

If $S(s,t)$ has stationary increments, the SSSC can be expressed as \cite{jiang2010note}
\begin{eqnarray}
\beta^{d}(t-s) &=& \frac{1}{-\theta}\log{\overline{M}_{S(s,t)}^{d}(\theta)},
\end{eqnarray}
with bounding function $g(x)=e^{-\theta x}$.

More generally, if $C(i)$ and $C(j)$, $i\neq j$, are dependent, 
replacing $r=\beta(t-s)-x$ in (\ref{dtccdf}) and (\ref{dtccdfl}), the SSSC of the capacity is obtained
\begin{equation}
g^{dl}_{s,t}(x)\equiv F_{S(s,t)}^{l}(\beta(t-s)-x) 
\le P\left\{S(s,t)\leq{\beta(t-s)-x}\right\}\le 
F_{S(s,t)}^{u}(\beta(t-s)-x) \equiv g^{du}_{s,t}(x).
\end{equation}

\subsubsection{Rayleigh channel as an example}

For the i.i.d. case, applying (\ref{istationary}),
\begin{IEEEeqnarray}{rCl}
{\beta^{ii}(t-s) = \frac{t-s}{-\theta}\log{\overline{M}_{C}(\theta) }} 
= \frac{t-s}{-\theta}\log\left( \frac{\ln{2}}{\gamma} \int_{0}^{\infty} {e^{-\theta{r}}} 2^{r}e^{-(2^r-1)/\gamma} d r \right),
\end{IEEEeqnarray}
with bounding function $g(x)=e^{-\theta x}$.

If $S(s,t)$ has stationary increments, the lower bound and upper bound of the SSSC can be expressed as
\begin{eqnarray}
\beta^{dl}(\tau) &=& \frac{1}{-\theta}\log{\overline{M}_{S(\tau)}^{du}(\theta)} \\
&=& \frac{1}{-\theta}\log \Big(  \tau(1-\exp(({1-2^{\Omega_{u}/{\tau}}})/{\gamma}-\theta{\Omega_{u}})) 
-\theta\tau \int_{0}^{\Omega_{u}} \exp\left( ({1-2^{r/\tau}})/{\gamma}-\theta{r}\right) d r  \Big), \\
\beta^{du}(\tau) &=& \frac{1}{-\theta}\log{\overline{M}_{S(\tau)}^{dl}(\theta)} \\
&=& \frac{1}{-\theta}\log \Big( \tau\exp\left(({1-2^{\Omega_{l}/\tau}})/{\gamma}-\theta{\Omega_{l}}\right)  
-\theta\tau \int_{\Omega_{l}}^{\infty} \exp\left( ({1-2^{r/\tau}})/{\gamma}-\theta{r}\right) d r  \Big),
\end{eqnarray}
where $\tau=t-s$, with bounding function $g(x)=e^{-\theta x}$.

For the dependent case, the lower bound and upper bound of the SSSC are expressed as
\begin{IEEEeqnarray}{rCl}
{P\left\{S(s,t)\leq{\beta(t-s)-x}\right\}} 
&\leq& \inf_{\sum\limits_{i=s+1}\limits^{t}r_{i}=\beta(t-s)-x}\left\{
\left[\sum\limits_{i=s+1}\limits^{t}{\left(1-e^{-(2^{r_{i}}-1)/{\gamma}}\right)} \right]_{1}\right\} \IEEEeqnarraynumspace\\
&\le& \left[ (t-s)\left( 1-\exp\left({-\left(2^{\frac{\beta(t-s)-x}{t-s}}-1\right)\Big{/}\gamma}\right) \right) \right]_{1} \equiv \dot{g}^{du}_{s,t}(x), 
 \IEEEeqnarraynumspace\\
{P\left\{S(s,t)\leq{\beta(t-s)-x}\right\}} 
&\ge& \!\!\sup_{\mbox{\scriptsize
$\begin{array}{c}
\sum\limits_{i=s+1}\limits^{t}r_{i} \\
=\beta(t-s)-x
\end{array}$}}\!\!\left[
\sum\limits_{i=s+1}\limits^{t}{\left(1-e^{-(2^{r_{i}}-1)/{\gamma}}\right)} -(t-s-1) 
\right]^{+} \IEEEeqnarraynumspace \\
&\ge&  \left[  1-(t-s)\exp\left({-\left(2^{\frac{\beta(t-s)-x}{t-s}}-1\right)\Big{/}\gamma}\right)   \right]^{+} \equiv \dot{g}^{dl}_{s,t}(x). \IEEEeqnarraynumspace
\end{IEEEeqnarray}

Let $\dot{g}^{du}_{s,t}(x)=\varepsilon$, $\tau=t-s$, and ${\beta}^{d}_{l}(\tau)=\beta(\tau)-x$, we obtain
\begin{equation}
{\beta}^{d}_l(\tau)=\tau\log_{2}\left( 1-\gamma\log\left( 1-\frac{\varepsilon}{\tau} \right) \right).
\end{equation}
Let $\dot{g}^{dl}_{s,t}(x)=\varepsilon$, $\tau=t-s$, and ${\beta}^{d}_{u}(\tau)=\beta(\tau)-x$, we obtain
\begin{equation}
{\beta}^{d}_u(\tau)=\tau\log_{2}\left( 1-\gamma\log\left( \frac{1-\varepsilon}{\tau} \right) \right).
\end{equation}

%
%
%

\section{Conclusion and Future Work}\label{conclusion}
The exploding data and increasing integration in future wireless communication put forth the exploration of the ultimate capacity that the wireless channel can provide while ensuring more and more stringent demand on quality of service. Novel information theoretic measures other than the conventional ergodic capacity or instantaneous capacity are required for network performance analysis. This paper has advocated a set of (new) concepts to study or as further properties of wireless channel capacity. They include cumulative capacity, maximum cumulative capacity, minimum cumulative capacity, and range of cumulative capacity, where for the latter three concepts, their forward-looking and backward-looking variations are also discussed. In addition, copula has been introduced as a technique to unify analysis of these introduced channel capacities under different dependence structures of the underlying cumulative channel capacity processes. Extensive results, including exact solutions for special cases and bounds for more general cases, for cumulative capacity are derived, mainly in terms of probability distribution function characteristics, e.g. CDF of the cumulative capacity process. In addition, its other characteristics in terms of moment generating function, Mellin transform, and stochastic service curve are also investigated. For maximum cumulative capacity and minimum cumulative capacity, preliminary study and results are presented. 

It is an on-going work is to extend the study on maximum and minimum cumulative capacity and use the range of cumulative capacity as a measure of tightness of the upper and lower bounds of the cumulative capacity. Both forward-looking and backward-looking variations of the general definitions are being studied. This work will be extended to apply the introduced concepts to QoS performance analysis of wireless channels, in addition to QoS-constrained capacities e.g. delay constrained capacity \cite{ephremides1998information,cai2006introduction,jiang2012stochastic}. At present, the analysis in this paper is based on a default assumption that the dependence structure of the cumulative service process is stationary over time. If the dependence behavior is time-variant, dynamic copula should be used to used for dependence modeling \cite{cherubini2011dynamic}, which will be our future work.

\section*{Acknowledgment}
\addcontentsline{toc}{section}{Acknowledgment}

The investigation on other characterizations, i.e., MGF, MT, and SSC, was started when Fengyou Sun was with the Department of Computer Science and Technology, Shanghai Normal University, China, under the supervision of Prof. Yuming Jiang and Prof. Luqun Li. Preliminary results of this investigation were included in \textit{Sun, Fengyou, Yuming Jiang, and Luqun Li. ``Further Properties of Wireless Channel Capacity.'' arXiv preprint arXiv:1502.00979v1 (2015).}

\bibliographystyle{IEEEtran}
\bibliography{main}

\end{document}